\documentclass[a4paper]{article}

\usepackage{INTERSPEECH2019}
\usepackage{caption}
\usepackage{subcaption}
\usepackage{siunitx}
\usepackage{multirow}

\usepackage[hidelinks]{hyperref}

\title{Analysis by Adversarial Synthesis - A Novel Approach for Speech Vocoding}
\name{Ahmed Mustafa$^*$\thanks{*The author was finishing his thesis with 1 and 2 during the writing of this paper. Currently, he is at Fraunhofer IIS in Erlangen (email: \textit{ahmed.mustafa.ahmed@iis.fraunhofer.de}).}, Arijit Biswas$^2$, Christian Bergler $^1$, Julia Schottenhamml$^1$, Andreas Maier$^1$}
\address{
  $^1$Pattern Recognition Lab, FAU Erlangen-Nuremberg, Erlangen, Germany\\
  $^2$Dolby Germany GmbH, Nuremberg, Germany}
\email{\{christian.bergler, julia.schottenhamml, andreas.maier\}@fau.de, arijit.biswas@dolby.com}

\begin{document}

\maketitle
\begin{abstract}
Classical parametric speech coding techniques provide a compact representation for speech signals. This affords a very low transmission rate but with a reduced perceptual quality of the reconstructed signals. Recently, autoregressive deep generative models such as WaveNet and SampleRNN have been used as speech vocoders to scale up the perceptual quality of the reconstructed signals without increasing the coding rate. However, such models suffer from a very slow signal generation mechanism due to their sample-by-sample modelling approach. In this work, we introduce a new methodology for neural speech vocoding based on generative adversarial networks (GANs). A fake speech signal is generated from a very compressed representation of the glottal excitation using conditional GANs as a deep generative model. This fake speech is then refined using the LPC parameters of the original speech signal to obtain a natural reconstruction. The reconstructed speech waveforms based on this approach show a higher perceptual quality than the classical vocoder counterparts according to subjective and objective evaluation scores for a dataset of 30 male and female speakers. Moreover, the usage of GANs enables to generate signals in one-shot compared to autoregressive generative models. This makes GANs promising for exploration to implement high-quality neural vocoders. 
\end{abstract}
\noindent\textbf{Index Terms}: speech coding, deep learning, generative adversarial networks, neural vocoder.

\section{Introduction}

    Speech coding is one of the fundamental functionalities of current multimedia communication systems over band limited transmission channels \cite{vary2006digital}. The conventional approaches for coding speech signals are based on the source-filter model, in which a speech signal is decomposed into its glottal excitation source signal and its vocal tract filter parameters \cite{vary2006digital}. Linear predictive coding (LPC) is used for implementing such source-filter modelling of speech signals \cite{vary2006digital}. This is incorporated in many speech coding standards such as AMR-WB \cite{bessette2002adaptive}. Vocoding is the process of describing speech signals in a fully parametric manner \cite{vary2006digital}. This provides the ability to build speech synthesizers operating based on the acoustic parameters which represent the target speech signal. The classical vocoding methods are developed based on hand-crafted acoustic parameters that mainly replace the glottal excitation, e.g. F0, VUV, etc. However, the reconstructed speech waveform according to such methods is well known to be synthetic and low in perceptual quality.
    
    Recently, autoregressive (AR) deep generative models have shown a great success in generating raw audio and speech waveforms especially after the emergence of WaveNet \cite{van2016wavenet} and SampleRNN \cite{mehri2016samplernn}. Both WaveNet and SampleRNN have been used as neural vocoders for reconstructing speech signals from the hand-crafted parametric representation of the source filter model \cite{kleijn2018wavenet, klejsa2019high}. The reconstructed speech signals using such neural vocoders are clearly high in their perceptual quality and even outperform the classical speech codecs such as AMR-WB. Unfortunately, this comes with the problem of very slow signal generation due to the sequential sampling of AR deep generative models. Therefore, additional techniques such as probability density distillation are required for running in real time \cite{pmlr-v80-oord18a}.

    Generative Adversarial Networks (GANs) provide an alternative approach for very fast generation of realistic data samples \cite{goodfellow2016nips}. GANs learn implicitly to estimate the PDF underlying the original data in order to directly generate new samples. This is achieved by a minimax adversarial training between a generator network that creates fake data and a discriminator network that compares it to the original data \cite{goodfellow2016nips}. When the training reaches an equilibrium state, the discriminator becomes fooled by the fake data created by the generator network, which is the target deep generative model.
    
    The usage of GANs for speech vocoding is very recent and was firstly introduced by Bollepalli et al. \cite{bollepalli2017generative} and Juvela et al. \cite{juvela2018speech}. The main idea of such approaches for GAN-based vocoding is to generate the glottal excitation signal adversarially and then apply synthesis filtering to obtain the speech waveform. However, this incorporates recurrent neural network architectures for predicting the voicing information and building a pulse model before creating the excitation signal with GANs.
    
    This paper proposes a new method for generating speech signal waveforms from a learned-compressed representation of the glottal excitation. The method uses GANs as an end-to-end fully-convolutional generative model that produces raw speech waveforms in one-shot. A simple refinement based on LPC is applied to the generated speech waveform in order to obtain a natural final reconstruction, without considerably affecting the overall complexity. An overview of this method is given in section 2, followed by a description of the experimental setup in section 3. Evaluation results are reported and discussed in section 4, to end up with conclusions in section 5.
\section{Analysis by Adversarial Synthesis}
Besides the ability of one-shot sample generation, GANs can create realistic data from a totally-abstract noise prior (e.g., Gaussian noise). The adversarial training makes it possible to map a simple prior distribution into complicated real-world distributions in a high-dimensional space. This has been achieved efficiently for image synthesis \cite{brock2018large} and recently for general audio synthesis \cite{engel2018gansynth}.

The speech vocoding task is described by a conditional generation process, so that the noise prior of the GAN model is converted into a parametric representation for the desired speech signal. To accomplish this, the glottal excitation signal, represented by the residual from an LPC analysis filtering of the speech waveform \cite{vary2006digital}, is fed to a neural encoder network. The residual is a noise-like signal as it is uncorrelated and almost spectrally-flat \cite{vary2006digital}. Thus, it is a good candidate to be compressed by the encoder network. This results in a learned conditional noise prior, which is a characteristic representative for the speech signal.
\subsection{Conditional Adversarial Synthesis}
Using the learned compressed representation of the residual signal as an input, a fake speech waveform is created using a deep generative model implemented by GANs. The generator network consists of two main blocks. The first block is called the \textit{context decoder}. It maps the conditional prior input (i.e., the context vector) into a multi-channel hidden representation to enable reliable signal generation. While the second block is called the \textit{adversarial upsampler} and it learns to upsample the context decoder output until reaching the desired signal resolution. The cascade of the residual neural encoder with the generator network gives a conditional generator model, which is trained jointly with the discriminator model.
\subsection{Cross Synthesis}
\begin{figure}
    \centering
    \includegraphics[height=3cm,width=\columnwidth]{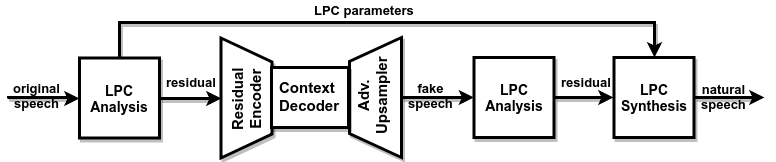}
    \caption{Illustration of AbAS.}
    \label{fig:abas}
\end{figure}

The fake speech waveform generated with GANs contains the main global and prosodic features of the target signal, especially at the first formants. However, some of phonemes and local details are missed due to the fast progressive generation of the speech waveform. Moreover, the adversarial training procedure does not follow a maximum-likelihood approach as in the AR generative models. This results in a considerable amount of reconstruction artifacts that affect the perceptual quality of the fake signal. To solve this issue, we propose to replace the spectral envelope of the fake speech with the original spectral envelope. This is done by an LPC analysis applied to the fake speech signal to obtain its fake residual. The fake residual is then filtered by the LPC parameters of the original speech signal to obtain a natural signal reconstruction. Hence, we name the whole process \textit{Analysis by Adversarial Synthesis} (AbAS), which is illustrated in Figure~\ref{fig:abas}.
\section{Model Configuration and Training Setup}
For training and testing the generative model, we used the clean speech signals of the dataset created by Valentini et al. \cite{valentini2016investigating}. It is an open source dataset of 15 male and 15 female speakers selected from the Voice Bank corpus introduced by Veaux et al. \cite{veaux2013voice}. The training data is constructed by the speech signals of 28 speakers, divided equally between males and females. While the test data are represented by speech of the remaining two speakers. The speech signals are downsampled from the original sampling rate of 48\,kHz to 16\,kHz which is our operating sampling rate. Furthermore, the corresponding glottal excitation signals are created by applying LPC analysis filtering of order 16 to the speech signals, with a frame length of 20\,ms.
\subsection{Residual Neural Encoder}
This network converts the LPC residual at sampling rate 16\,kHz into a learned context vector of 1\,kHz. The context vector is the conditional prior required for generating the target fake speech. The network consists of a stack of 4 downsampling convolutional layers. The downsampling is done by a 1D-convolution operation with kernel width 64 and stride 2, so that each layer downsamples its input by a factor of two. Parametric rectified linear unit (PReLU) \cite{he2015delving} is used for activation. Reflection padding is used for adjusting the signal length during the learned downsampling process. This results in the following feature maps starting from the input residual until the output of the fourth layer: 16000$\times$1, 8000$\times$32, 4000$\times$64, 2000$\times$64, 1000$\times$128. Finally, the output of the fourth layer is fed to a compressor represented by a convolutional layer with kernel size 65 and one output channel to obtain the context vector.
\subsection{Generator Network}
\subsubsection{Softmax-gated CNN}
One important feature of the generator network is the softmax-gated CNN layer. It is defined and implemented similarly to the sigmoid-gated CNN of WaveNet \cite{van2016wavenet}. However, the sigmoid operation is replaced by a softmax along the channel dimension of the gate output. Thus, the output of this gated layer is given as follows:
\begin{equation}
output = \tanh{(W_{f} * X)} \odot softmax(W_{g} * X)|_{c},
\end{equation}
where $X$ is the input to the gated-CNN layer, $W_{f}$ are the weights of the 1D-convolutional filter, $W_{g}$ are the weights of the 1D-convolutional gate, $*$ denotes the convolution operation and $\odot$ denotes the element-wise multiplication. For all gated-CNNs layers in the generator model, a kernel of width 65 is used for both the filtering and gating operations with reflection padding to maintain the signal length.
%\begin{figure}
%    \centering
%    \includegraphics[height=5cm,width=0.6\linewidth]{softmax_gate_updated.png}
%    \caption{The softmax-gated CNN layer.}
%    %\label{fig:abas}
%\end{figure}
\subsubsection{Context Decoder}
This block consists of a stack of 10 identical gated-CNN layers that generate a hidden representation of 64 channels for the context vector learned by the residual encoder. It was found more effective than direct upsampling as it reduces the reconstruction artifacts of the generated fake signal. A 1$\times$1 convolution operation precedes this block to create 64 channels of the context vector ready for manipulation.
\subsubsection{Adversarial Upsampler}
\begin{figure}
    \centering
    \includegraphics[height=7cm,width=\columnwidth]{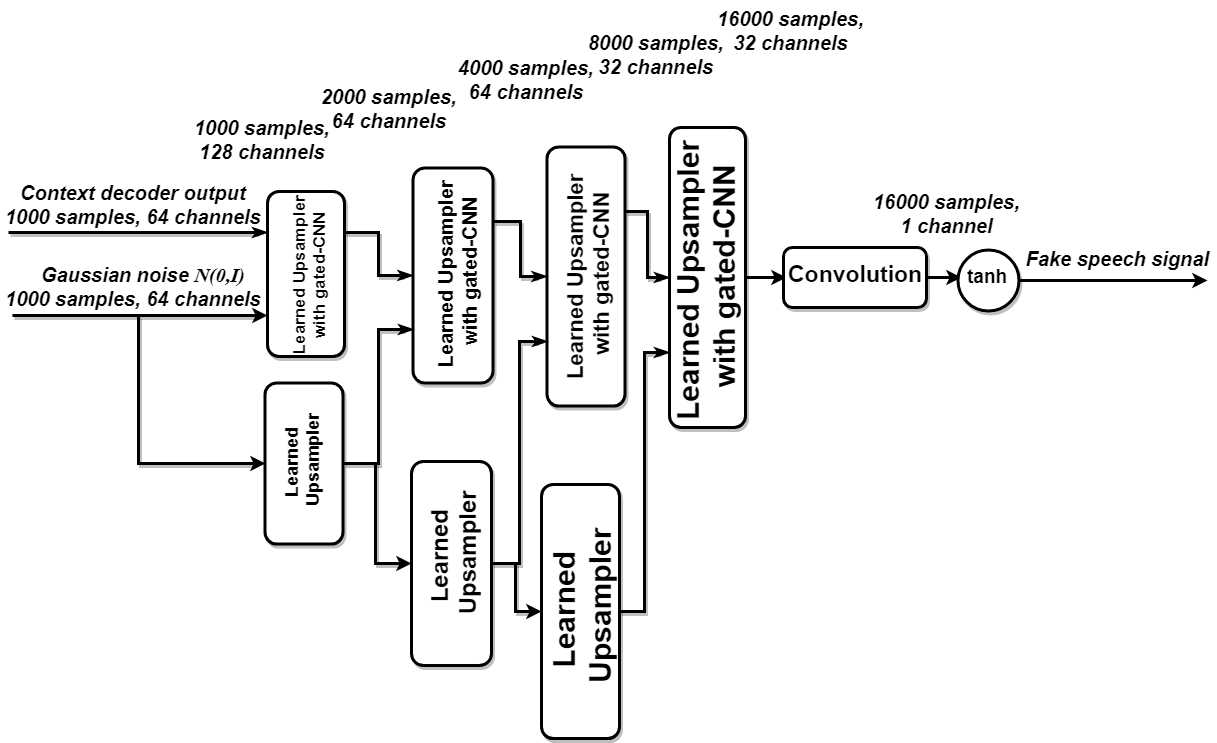}
    \caption{The adversarial upsampler network.}
    \label{fig:upsampler}
\end{figure}

The adversarial upsampler converts the multi-channel context decoder output of 1\,kHz/channel into a single channel fake speech of 16\,kHz. This is done by progressive upsampling using 4 layers. Each layer applies 1D-transposed convolution with kernel width 66 and stride 2 in order to obtain an output with doubled length compared to the layer input. Moreover, each layer passes the output of the transposed convolution through a gated-CNN without changing the dimensionality to refine and activate the upsampling. In parallel to this, a Gaussian noise of zero mean and unit variance is independently upsampled and shaped using transposed convolution without activation. This noise is used for compensating the missing fine details of the speech signal during the residual compression task, e.g. unvoiced speech parts and background noise. It is then concatenated along the channel dimension with the actual signal generation path at every upsampling stage. The upsampler block diagram and the feature maps throughout the signal generation path are illustrated in Figure~\ref{fig:upsampler}.
Note that the Gaussian noise has the same dimensionality as the signal feature maps at every upsampling stage. So that the input channels at each signal upsampling stage are equally divided between the noise channels and the signal channels from the previous stage. The output layer applies 1D-convolution with kernel width 65 and tanh non-linearity. 
\subsection{Conditional Adversarial Training}
A conditional generative adversarial network (CGAN) is trained with the same technique used for image-to-image translation \cite{isola2017image}. The conditional discriminator $D$ receives a 2-channel concatenation of the residual and the corresponding original/fake speech signals. The network of $D$ comprises 6 1D-CNN layers with stride 2 and kernel width 32 per each. LeakyReLU \cite{he2015delving} with a leakage factor of 0.2 is used for activating all layers, except the last one where only the convolution operation is applied. The channel depths starting from the input until the output of $D$ are: 2, 16, 16, 32, 32, 64 and 32. Spectral normalization \cite{miyato2018spectral} is applied to all convolutional layers of $D$ to ensure the Lipschitz continuity that is required for stable adversarial training using distance-based loss functions \cite{arjovsky2017wasserstein}. The conditional generator $G$ is the cascade of the residual encoder, the context decoder and the adversarial upsampler networks which are trained jointly. We have also applied spectral normalization to all convolutional layers of $G$ as this was found helpful for better training stability \cite{zhang2018self}. The training of $D$ is driven by the adversarial hinge loss \cite{zhang2018self}:
\begin{equation}
\begin{split}
L_D = -\mathbb{E}_{x \sim p_{speech}, r \sim p_{residual}} [\min(0, -1 + D(x,r))] \\
-\mathbb{E}_{z \sim p_{z}, r \sim p_{residual}} [\min(0, -1 - D(G(z,r),r))],
\end{split}
\end{equation}
where $L_D$ is the total conditional loss of $D$, $p_{speech}$ denotes the original speech data, $p_{residual}$ denotes the residual data, $p_{z}$ denotes the Gaussian noise used during the adversarial upsampling and $G(z,r)$ denotes the fake speech data. For training $G$, the total loss function $L_G$ is given by the following convex form:
\begin{equation}
\begin{split}
L_G = \gamma\mathbb{E}_{x \sim p_{speech}, z \sim p_{z}, r \sim p_{residual}}[||G(z,r) - x ||_{1}] \\ -(1-\gamma) \mathbb{E}_{z \sim p_{z}, r \sim p_{residual}}[D(G(z,r),r)],
\end{split}
\end{equation}
with regularization factor $\gamma$ of 0.00015. Both $D$ and $G$ are optimized alternately with equal number of training iterations. Adam optimizer with AMSG\textsubscript{RAD} \cite{DBLP:conf/iclr/ReddiKK18} is used, with lr\,=\,0.0006 for $D$ / 0.00015 for $G$ and $\beta$\,=\,[0.5, 0.99]. Xavier algorithm \cite{glorot2010understanding} is used for initializing the weights for both $D$ and $G$. The batch size is 32. 
\section{Results}
\begin{figure}[!htbp]
  \centering
  \begin{subfigure}[c]{\columnwidth}
    \includegraphics[height=2.5cm,width=\columnwidth]{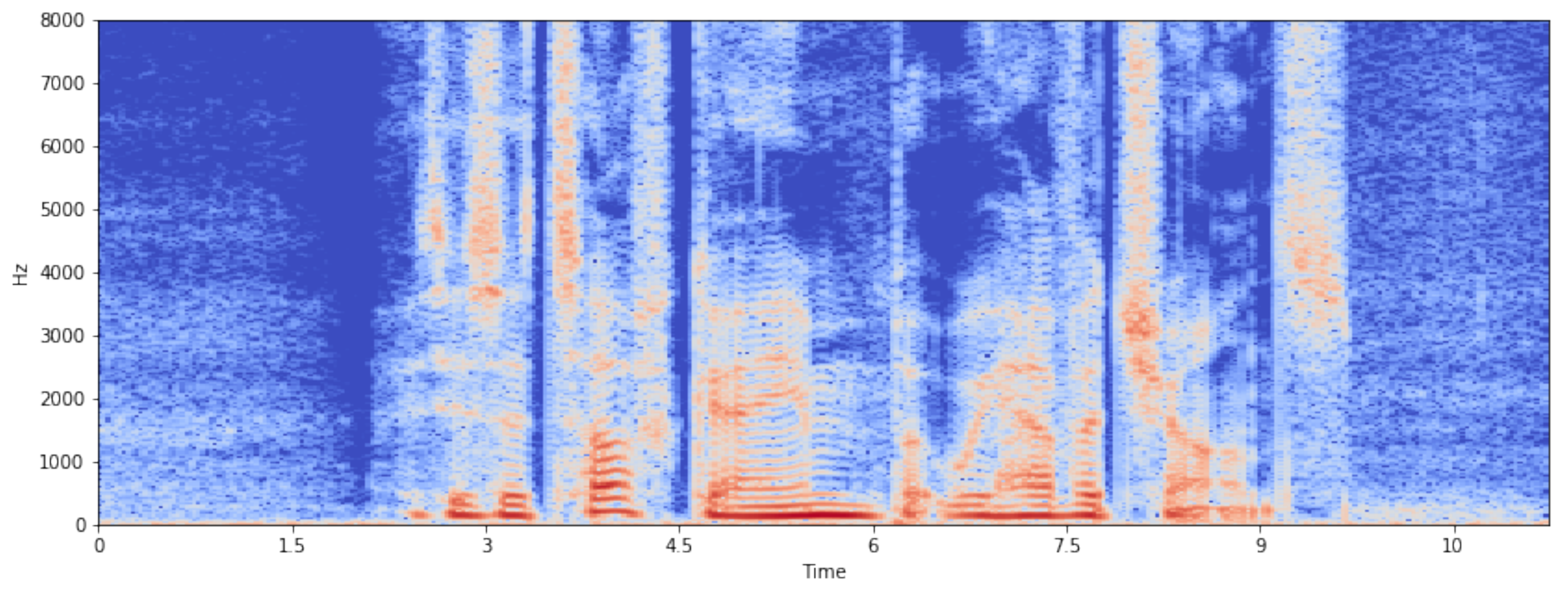}
    %\subcaption{Original speech signal.}
    %\label{fig:f1}
  \end{subfigure}
  %\vfill
  \begin{subfigure}[c]{\columnwidth}
    \includegraphics[height=2.5cm,width=\columnwidth]{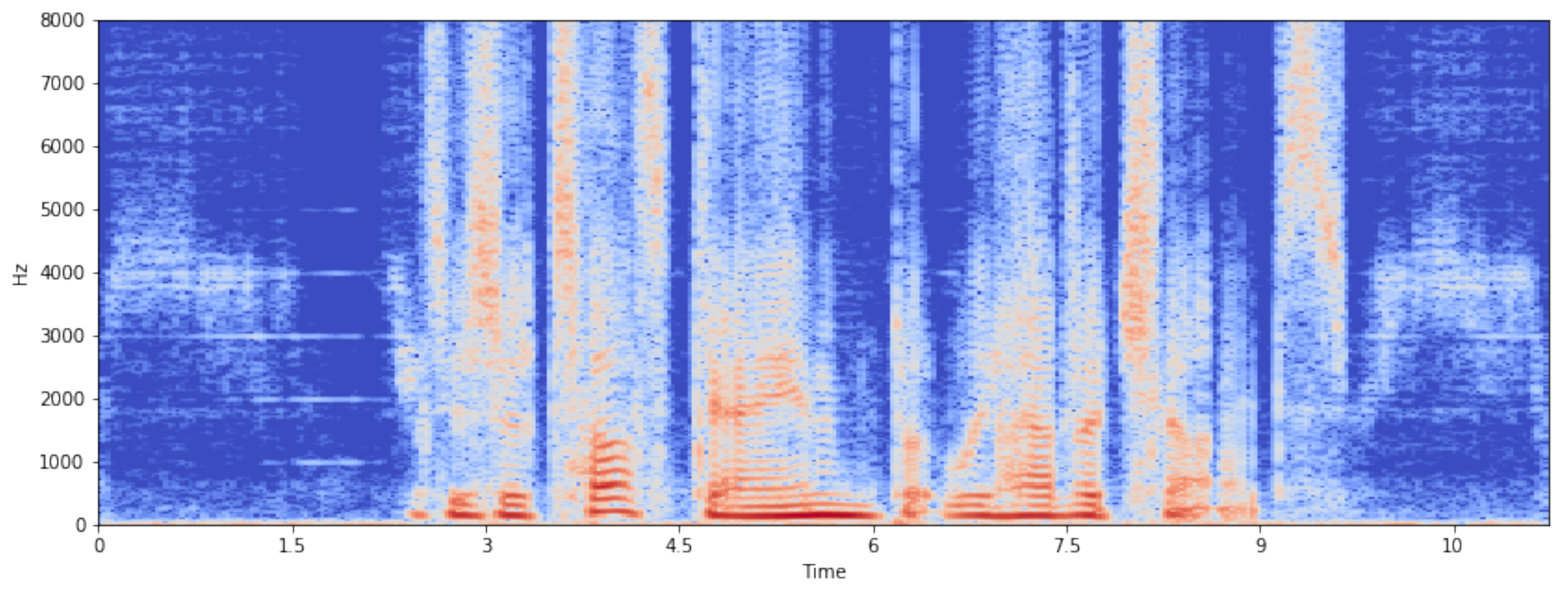}
    %\subcaption{Fake speech signal generated from the neurally-compressed residual.}
    %\label{fig:res2fake}
  \end{subfigure}
   %\vfill
  \begin{subfigure}[c]{\columnwidth}
    \includegraphics[height=2.5cm,width=\columnwidth]{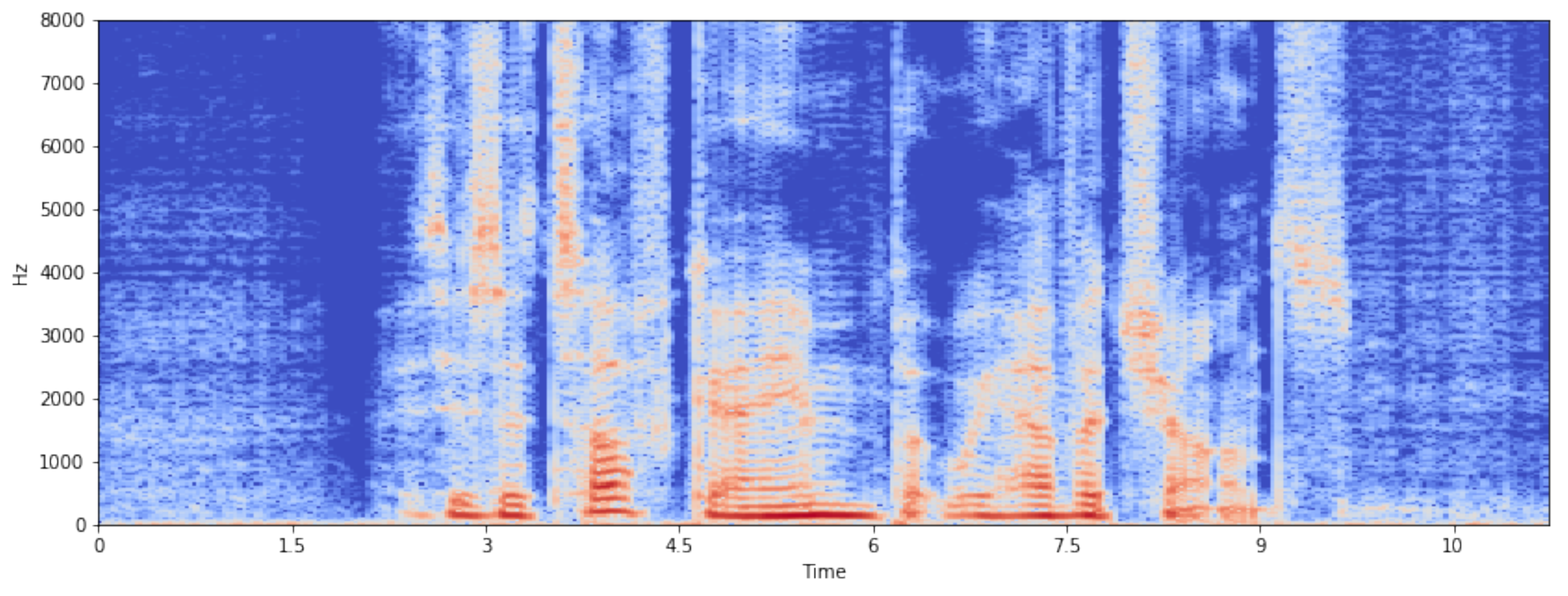}
    %\subcaption{Natural speech signal reconstructed with AbAS}
    %\label{fig:abas_recon}
  \end{subfigure}
    \caption{GANs for speech vocoding: A fake speech signal is generated by CGAN (middle) at 16\,kHz from the 1\,kHz learned compression of the residual signal. This fake signal preserves the main spectral and prosodic features of the original speech (top) especially at the low frequency bands. However, it is more challenging to accurately reconstruct the high frequency details and the background noise of the original signal. That is why the cross synthesis step is incorporated to refine the fake speech signal using the original LPC parameters, which results in a natural final reconstruction (bottom).}
  \label{fig:gan_vocodig}
\end{figure}

The main outcome of this work is the ability of CGANs to create realistic speech waveforms in one-shot from a highly compressed representation of the glottal excitation. This is enhanced by the cross synthesis step in order to obtain a natural reconstruction, as illustrated in Figure~\ref{fig:gan_vocodig}.
%\begin{figure}[!htbp]
%    \centering
%    \includegraphics[height=2.2cm,width=\columnwidth]{fake_no_gate.pdf}
%    \caption{Fake speech using PReLU activation in the generator network.}
%    \label{fig:res2sig_no_gate}
%\end{figure}

\begin{figure}[!htbp]
    \centering
    \includegraphics[height=3.5cm,width=\columnwidth]{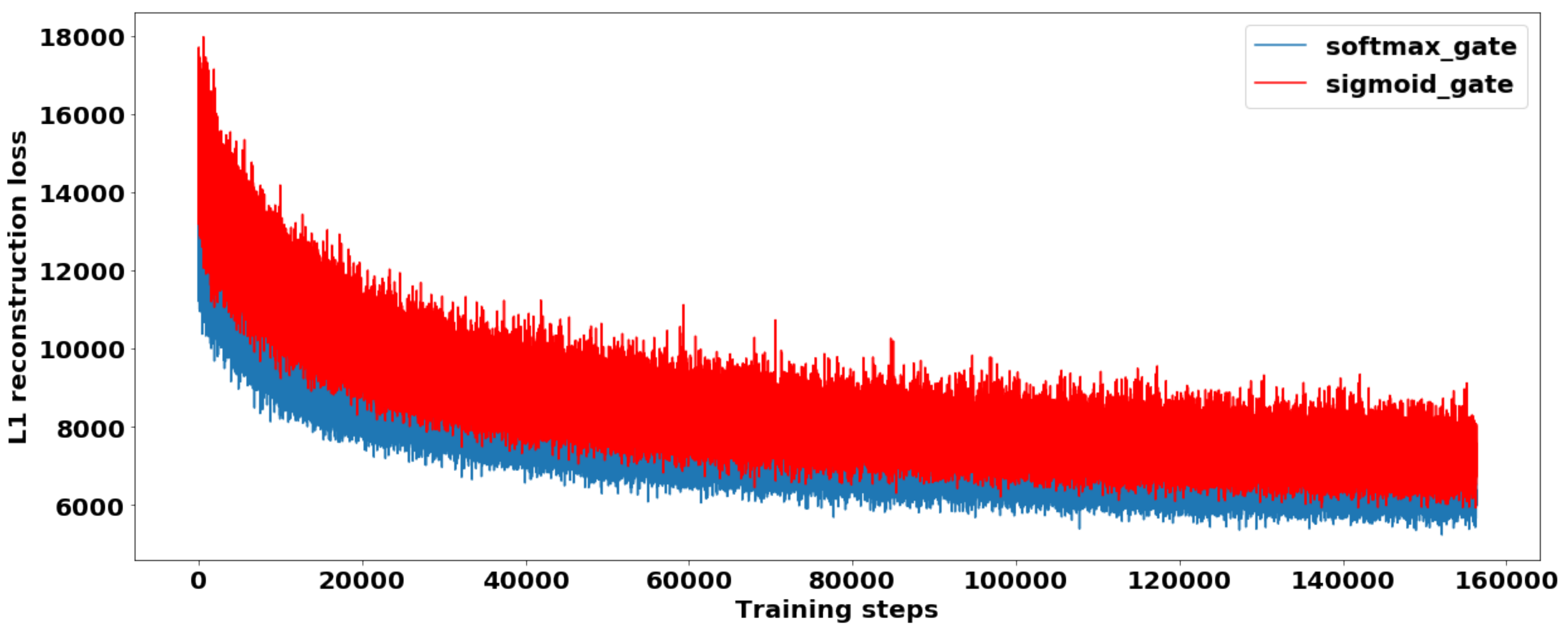}
    \caption{Outperformance of the proposed softmax gating over the sigmoid one in terms of the L1 reconstruction loss.}
    \label{fig:softmax_vs_sigmoid}
\end{figure}

The gated activation was found more robust than the ReLU-based one for penalizing the reconstruction artifacts during the signal generation. Furthermore, using the softmax along the channel dimension of feature maps of the gate output is more effective than the element-wise sigmoid. The L1 loss curve of the conditional generator shows a faster decay with softmax than the sigmoid case, as illustrated by Figure~\ref{fig:softmax_vs_sigmoid}. A possible reason for this is that the softmax along the channel dimension models the relationship between the frequency bins of the signal at every time instant. This leads to a probability mask that gives a higher weight to the frequency bins which are more relevant to the desired samples of the target signal, while penalizing the artifacts with lower weights to their frequency components.

The proposed AbAS approach is assessed by objective and subjective perceptual evaluation measures. This is done in comparison with the classical vocoder introduced by Hedelin \cite{hedelin2000sinusoidal} and refined by Klejsa et al. \cite{klejsa2019high}. There is no quantization applied to the compressed representation of signals for both the classical vocoder and AbAS. We only focus on evaluating the signals reconstructed from the non-quantized parametric representation.
\subsection{Objective Evaluation}
We resort to the 5 objective measures used by Pascual et al. for evaluating the SEGAN \cite{Pascual2017SEGANSE}: PESQ-WB, CSIG, CBAK, COVL, SSNR. In addition, we use the ViSQOL perceptual objective score \cite{Hines2015}. All of these measures give their results in mean opinion score (MOS), except for the SSNR which is given in dB. This ensures a precise evaluation of the proposed approach in terms of perceptual quality and robustness against the reconstruction artifacts. Table ~\ref{tab:objective_measures} shows how AbAS outperforms the classical vocoder reconstructions.
\begin{table}[th]
  \caption{Objective evaluation results, 40 speech signals are randomly selected from both the training and testing datasets for assessment. }
  \label{tab:objective_measures}
  \centering
  \begingroup
  \setlength{\tabcolsep}{4.5pt} % Default value: 6pt
  \renewcommand{\arraystretch}{1.1} % Default value: 1
  \begin{tabular}{c cc cc}
    \toprule
    \multirow{2}{*}{\textbf{Metric}} &
      \multicolumn{2}{c}{\textbf{ValidationSet}} &
      \multicolumn{2}{c}{\textbf{TestSet}} \\
      & {\textit{Vocoder}} & {\textit{AbAS}} & {\textit{Vocoder}} & {\textit{AbAS}} \\
      \midrule
    ViSQOL[MOS] & 2.9062 & 3.1355 & 2.9161 & 3.1065 \\
    
    PESQ-WB[MOS] & 2.5138 & 2.8234 & 2.3065 & 2.5977 \\
    
    CSIG[MOS] & 4.1234 & 4.4876 & 3.9771 & 4.3143 \\
    
    CBAK[MOS] & 2.4697 & 3.0317 & 2.3760 & 2.8720 \\
    
    COVL[MOS] & 3.2941 & 3.6596 & 3.1146 & 3.4534 \\
    
    SSNR[dB] & -2.1973 & 3.1438 & -2.0200 & 2.5865 \\
    \bottomrule
  \end{tabular}
  \endgroup
  
\end{table}
\subsection{Subjective Evaluation}
A MUSHRA listening test \cite{itu20031534} is performed by 7 subjects to evaluate the perceptual quality of 20 reconstructed speech signals for males and females. The CGAN model is trained with 600 epochs. Most of the AbAS reconstructions are more perceptually preferred than the classical vocoder ones. It is worth mentioning that longer training with more data should give better results due to the better approximation of data modalities by CGAN. But we just emphasize here the proof of the concept. It was also found that the perceptual quality can be scaled by increasing the cross synthesis parameters as this compensates the degradation of the missing phonemes which are not well reconstructed with CGAN due to the sub-optimal distribution modelling. 
\begin{figure}[!htbp]
    \centering
    \includegraphics[height=3.5cm,width=\columnwidth]{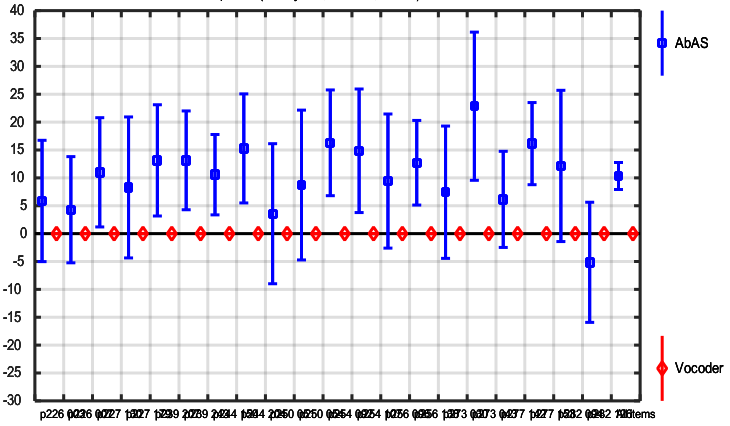}
    \caption{MUSHRA differential scores of AbAS w.r.t classical vocoder.}
    \label{fig:mushra}
\end{figure}
%\subsection{Setting of the LPC Filter Order}
%\begin{figure}[!htbp]
%    \centering
%    \includegraphics[height=2.5cm,width=0.5\textwidth]{cs_quality_scale.pdf}
%    \caption{Quality scaling with increasing LPC cross synthesis parameters.}
%    \label{fig:cs_quality_scale}
%\end{figure}

\begin{figure}[!htbp]
    \centering
    \includegraphics[height=3.5cm,width=\columnwidth]{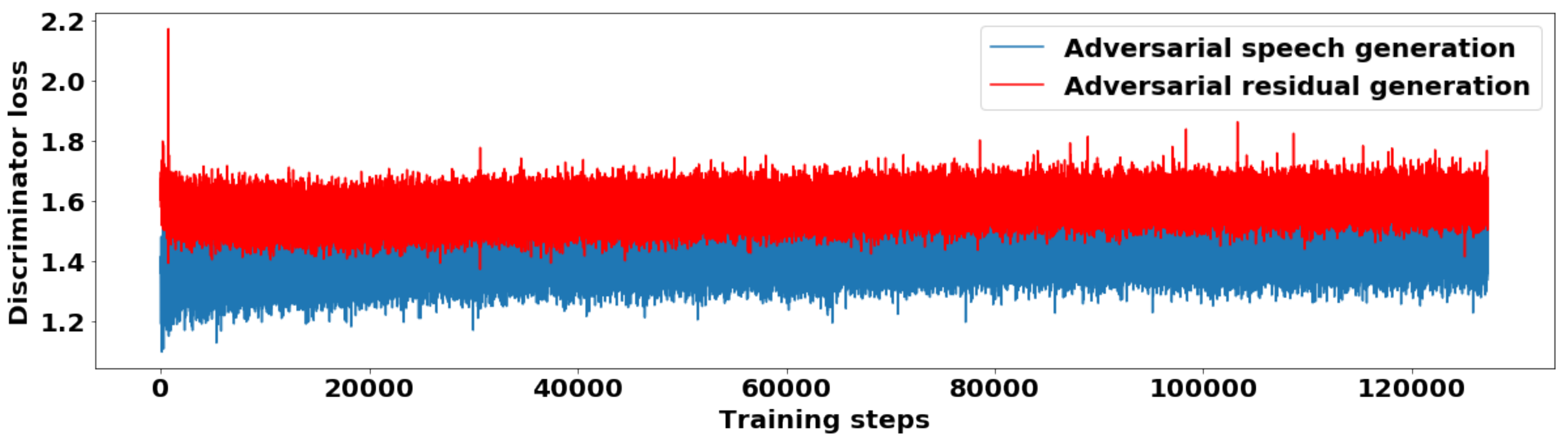}
    \caption{Higher discriminator loss for generating fake residual compared to fake speech, which indicates a lower quality for the generated residual samples.}
    \label{fig:disc_res_speech}
\end{figure}

Instead of AbAS, we tried to generate a fake residual from a very compressed representation with CGAN and hence apply an LPC synthesis using the original LPC parameters to reconstruct the speech signal. However, this gave poor results compared to AbAS. That is because the discriminator is stronger in rejecting fake uncorrelated signals (i.e., residuals) than correlated ones, which makes it harder to generate realistic residuals from a compressed representation. Figure.~\ref{fig:disc_res_speech} illustrates this finding.
\section{Conclusions}
This paper introduces a new method for neural speech vocoding, with much faster generation than autoregressive generative models and higher perceptual quality than classical vocoding. The method, which is called analysis by adversarial synthesis (AbAS), starts with generating a fake speech signal from a neurally-learned parametric representation of the glottal excitation using conditional GANs. This is accompanied by an LPC cross synthesis step, using the spectral envelope parameters of the original speech, to obtain a natural reconstruction. A possible future work is to explore better convolutional architectures for the generator model to reduce the reconstruction artifacts. Further work can be done to investigate the possibility of predicting the cross synthesis parameters from the fake speech. This makes it promising to optimize this approach for competing with advanced classical speech codecs at considerably lower coding rates. 

\bibliographystyle{IEEEtran}
\bibliography{mybib}

\end{document}